\documentclass[12pt, onecolumn]{IEEEtran}
\usepackage{epsf}
\usepackage{graphicx}
\usepackage{amsmath} \usepackage{amssymb}

\newtheorem{theorem}{Theorem}[section]

\long\def\symbolfootnote[#1]#2{\begingroup%
\def\thefootnote{\fnsymbol{footnote}}\footnote[#1]{#2}\endgroup}

  \def\C{\mathbb C} \def\RR{{\cal R}}

  \def\C{\Sigma}

\def\dref#1{(\ref{#1})}

\def\be{\begin{equation}} \def\ee{\end{equation}}
\def\ba{\begin{array}} \def\ea{\end{array}} \def\bna{\begin{eqnarray}}
\def\ena{\end{eqnarray}}

 \def\NN{{\cal N}}

\def\C{\mathbb C}

\def\SS{{\cal S}}
 \def\XX{{\cal X}}   
  
\def\TT{{\cal T}}

\def\YY{{\cal Y}}

 \def\bna{\begin{eqnarray}}
\def\ena{\end{eqnarray}} \def\dref#1{(\ref{#1})}
\begin{document}
\title{A Greedy Omnidirectional Relay Scheme}

\author{\authorblockN{Liang-Liang Xie} \\
\authorblockA{\small Department of Electrical and Computer Engineering\\
University of Waterloo, Waterloo, ON, Canada N2L 3G1 \\
Email: llxie@ece.uwaterloo.ca} }

\maketitle

\begin{abstract}

A greedy omnidirectional relay scheme is developed, and the corresponding achievable rate region is obtained for the all-source all-cast problem. The discussions are first based on the general discrete memoryless channel model, and then applied to the additive white Gaussian noise (AWGN) models, with both full-duplex and half-duplex modes.

\end{abstract}

\section{Introduction}

A general framework of omnidirectional relay has been developed in \cite{xie07a}-\cite{xie08b}. It generalizes the decode-and-forward relay strategy introduced in \cite{covelg79} with the network coding idea introduced in \cite{ahlcailiyeu00} to the case of wireless networks with multiple sources. Technically, it is a combination of block Markov coding with binning, so that each relay can simultaneously transport multiple messages in different directions. The effectiveness of this omnidirectional relay strategy has been demonstrated by the result that it is possible to completely eliminate interference in the network, and each node can fully exploit the signals transmitted by all the other nodes.

In this paper, we develop a special ``greedy'' omnidirectional relay scheme in the sense that each node tries to relay as many messages as possible. Without being regulated by network topologies, this greedy scheme is simple to implement, and can be adaptive to time-varying situations.

Our discussion will first be on the general discrete memoryless channel model. And then, motivated by wireless networks, the results will be applied to the AWGN models, with both full-duplex and half-duplex modes. For simplicity, in this paper, we focus on the all-source all-cast problem, and obtain a general achievable rate region.

\section{A General Discrete Memoryless Network Channel Model}

Consider a network of $n$ nodes $\NN=\{1,2,\ldots,n\}$, with the channel modeled by
$$
(\XX_1\times \cdots \times \XX_n, p(y_1,\ldots,y_n|x_1,\ldots,x_n), \YY_1\times \cdots \times \YY_n).
$$
At each time $t=1,2,\ldots$, every node $i\in \NN$ sends an input $X_i(t)\in \XX_i$, and receives an output $Y_i(t)\in\YY_i$, and they are related via $p(Y_1(t),\ldots,Y_n(t)|X_1(t),\ldots,X_n(t))$.

\section{A Greedy Omnidirectional Relay Scheme}
\label{scheme}

The essence of this ``greedy'' scheme is that at the end of each block, every node decodes as many messages as possible, and in the next block, relays all the messages it has decoded, with the restriction of adding at most one new message for each source. To be more specific, every node $i$ relays the message $w_j(b_0)$, if it has decoded it, and it has relayed all the messages $w_j(b),\,b=1,\ldots,b_0-1$ previously.

Consider the all-source all-cast problem, where each node $i$ is an independent source, and wants to send some common information to all the other nodes at the rate $R_i$. With this greedy omnidirectional relay scheme, we have the following achievable rate region for the all-source all-cast problem.

\begin{theorem}
\label{th1}
Consider the all-source all-cast problem. With the greedy omnidirectional relay scheme, a rate vector $(R_1,R_2,\ldots, R_n)$ is achievable if for any nonempty subset $\SS\subset \NN$, there is a node $i_0\in \SS$, such that
\begin{equation}
\label{eq1}
\sum_{j\in\SS^c}R_j<I(X_{\SS^c};Y_{i_0}|X_\SS)
\end{equation}
for some $p(x_1)p(x_2)\cdots p(x_n)$, where $X_{\SS^c}=\{X_j:j\in\SS^c\}$, and $X_{\SS}=\{X_i:i\in\SS\}$.
\end{theorem}

For three-node networks, the achievability of the rate region prescribed by \dref{eq1} has been proved in \cite[Thm 4.1]{xie07}, where, instead of the greedy relay scheme, the relay ordering was set according to the relative strengths of the channels between different nodes. However, even for three-node networks, the proof in \cite{xie07} turned out to be rather complicated, since there were too many different cases to address. Here, in Section \ref{proof} of this paper, we will present a simple and general proof based on the greedy relay scheme, which applies to networks with any number of nodes.

Now, we consider a time-varying operation of the network, with different input distributions in different blocks. Specially, we are interested in the periodic case, where the input distribution in block $b$ is $p_k(x_1)p_k(x_2)\cdots p_k(x_n)$ with $k=(b \mbox{ mod } K)$ for some period  $K\geq 2$. Correspondingly, we have the following conclusion.

\begin{theorem}
\label{th2}
Consider the all-source all-cast problem. With the periodic greedy omnidirectional relay scheme, a rate vector $(R_1,R_2,\ldots, R_n)$ is achievable if for any nonempty subset $\SS\subset \NN$, there is a node $i_0\in \SS$, such that
$$
\sum_{j\in\SS^c}R_j<\frac{1}{K}\sum_{k=1}^K I_k(X_{\SS^c};Y_{i_0}|X_\SS)
$$
where, the mutual information $I_k$ is calculated based on $p_k(x_1)p_k(x_2)\cdots p_k(x_n)$.
\end{theorem}

Obviously, to obtain more general results, we can also consider different block lengths. Let block $b$ have length $L_k$ with $k=(b \mbox{ mod } K)$. Then, we have the following conclusion.

\begin{theorem}
\label{th3}
Consider the all-source all-cast problem. With the periodic greedy omnidirectional relay scheme with varying block lengths, a rate vector $(R_1,R_2,\ldots, R_n)$ is achievable if for any nonempty subset $\SS\subset \NN$, there is a node $i_0\in \SS$, such that
$$
\sum_{j\in\SS^c}R_j<\frac{1}{\sum_{k=1}^K L_k}{\sum_{k=1}^K L_kI_k(X_{\SS^c};Y_{i_0}|X_\SS)}
$$
where, the mutual information $I_k(\cdot)$ is calculated based on $p_k(x_1)p_k(x_2)\cdots p_k(x_n)$.
\end{theorem}

\section{Full-Duplex AWGN Wireless Networks}

Consider the following AWGN wireless network
channel model with full-duplex mode:
\begin{equation}\label{netcha1}
Y_j(t)=\sum_{\stackrel{i\in \NN}{i\neq
j}}g_{i,j}X_i(t)+Z_j(t),\quad \quad \forall\, j\in \NN,\quad
t=1,2,\ldots
\end{equation}
where, $X_i(t)\in \C^1$ and $Y_i(t)\in\C^1$ respectively denote
the signals sent and received by Node $i\in\NN$ at time $t$; $\{ g_{i,j} \in
\C^1 : i \neq j \}$ denote the signal attenuation gains; and
$Z_i(t)$ is zero-mean complex Gaussian noise with variance $N$.

Consider the average power constraint:
$$
\frac{1}{T}\sum_{t=1}^{T}|X_i(t)|^2 \leq P_i \quad \mbox{ for all }T=1,2,\ldots,\mbox{ and }  i\in \NN.
$$
Then applying Theorem \ref{th1}, we have the following conclusion.
\begin{theorem}
\label{th4}
Consider the all-source all-cast problem for the full-duplex AWGN wireless networks. With the greedy omnidirectional relay scheme, a rate vector $(R_1,R_2,\ldots, R_n)$ is achievable if for any nonempty subset $\SS\subset \NN$, there is a node $i_0\in \SS$, such that
$$
\sum_{j\in\SS^c}R_j<\log\left(1+\frac{\sum_{j\in \SS^c}|g_{j,i_0}|^2P_j}{N}\right).
$$
\end{theorem}

\section{Half-Duplex AWGN Wireless Networks}

Consider the following AWGN wireless network
channel model with half-duplex mode: At time $t=1,2,\ldots$, the transmitter set is $\TT(t)\subset \NN$, and the receiver set is $\RR(t)=\NN\backslash \TT(t)$, and
\begin{equation}\label{netcha2}
Y_j(t)=\sum_{{i\in \TT(t)}}g_{i,j}X_i(t)+Z_j(t),\quad \quad \forall\, j\in \RR(t),
\end{equation}
where, $X_i(t)\in \C^1$ and $Y_j(t)\in\C^1$ respectively denote
the signal sent by node $i$ and the signal received by node $j$ at time $t$; $\{ g_{i,j} \in
\C^1 : i \neq j \}$ denote the signal attenuation gains; and
$Z_j(t)$ is zero-mean complex Gaussian noise with variance $N$.

Consider the following average power constraint:
$$
\frac{\sum_{t=1}^{T}|X_i(t)|^2{\mathbb{I}_{[i\in\TT(t)]}}}{\sum_{t=1}^T {\mathbb{I}_{[i\in\TT(t)]}}} \leq P_i \quad \mbox{ for all }T=1,2,\ldots,\mbox{ and }  i\in \NN,
$$
where ${\mathbb{I}_{[\cdot]}}$ is the indicator function:
$$
{\mathbb{I}_{[i\in\TT(t)]}}=\left\{ \begin{array}{ll} 1, & \mbox{ if } i\in\TT(t), \\ 0, & \mbox{ otherwise.}\end{array}\right.
$$

Consider a periodically block-varying operation of the network. In block $b=1,2,\ldots$, the block length is $L_{k}$, the transmitter set is $\TT_k$, and the receiver set is $\RR_k$, with $k=(b \mbox{ mod } K)$ for some period $K\geq 2$. Then by Theorem \ref{th3}, we have the following conclusion.
\begin{theorem}
\label{th5}
Consider the all-source all-cast problem for the half-duplex AWGN wireless networks. With the periodic greedy omnidirectional relay scheme with varying block lengths, a rate vector $(R_1,R_2,\ldots, R_n)$ is achievable if for any nonempty subset $\SS\subset \NN$, there is a node $i_0\in \SS$, such that
$$
\sum_{j\in\SS^c}R_j<\frac{1}{\sum_{k=1}^K L_k}\sum_{k=1}^KL_k{\mathbb{I} _{[i_0\in \RR_k]}}\log\left(1+\frac{\sum_{j\in \SS^c\cap \TT_k}|g_{j,i_0}|^2P_j}{N}\right).
$$
\end{theorem}

\section{Proof of the Theorems}
\label{proof}

\paragraph*{Proof of Theorem \ref{th1}} The key to the proof is the technical Lemma 4.1 developed in \cite{xie08b}, which basically says that once the inequality \dref{eq1} holds, node $i_0$ can always decode the messages of some nonempty subset of $\SS^c$. We will prove by induction that each node can decode the messages sent by all the other nodes.

According to the greedy relay scheme, once a node $i$ has decoded some messages of another node $j$, it will always transmit the messages of node $j$ in the subsequent blocks. We say that node $i$ covers a set of nodes $\SS$, if node $i$ has decoded some messages of every node in $\SS$, and therefore, will transmit the messages of every node in $\SS$ in the subsequent blocks. It is obvious that at the end of any block $b\geq 1$, each node $i$ can decode the block-$b$ transmission of some other node $j_i\neq i$, by applying the Lemma to \dref{eq1} with $\SS=\{i\}$. In other words, at the end of block $b$, each node $i$ will at least cover what have been covered by certain two nodes $\{j_i,i\}$ at the end of block $b-1$. For $b\geq 2$, since at the end of block $b-1$, each one of the two nodes $\{j_i,i\}$ must have covered what had been covered by at least a pair of nodes at the end of block $b-2$, we have that at the end of block $b$, node $i$ will at least cover what had been covered by three nodes at the end of block $b-2$. To see this, there are two cases: If at least one of the two pairs is different from $\{j_i,i\}$, the total covering is obviously at least three nodes; If both the two pairs are identical to $\{j_i,i\}$, then one of the two nodes $\{j_i,i\}$ must be able to cover another node according to the Lemma applied to \dref{eq1} with  $\SS$ set to $\{j_i,i\}$, thus still leading to a covering of at least three nodes. Therefore, at the end of any block $b\geq 2$, each node will at least cover what had been covered by certain three nodes at the end of block $b-2$.

Now, since at the end of any block $b\geq 4$, each node $i$ at least covers what had been covered by certain three nodes $\{j_i, k_i, i\}$ at the end of block $b-2$, while each of them in turn must have covered what had been covered by a set of three nodes at the end of block $b-4$, we have that at the end of block $b$, node $i$ will at least cover what had been covered by four nodes at the end of block $b-4$. To see this, similarly there are two cases: If at least one of the three sets is different from $\{j_i, k_i, i\}$, the total covering is at least four nodes; If all the three sets are identical to  $\{j_i, k_i, i\}$, then one of the three nodes  $\{j_i, k_i, i\}$ must be able to cover another node according to the Lemma applied to \dref{eq1} with  $\SS$ set to $\{j_i,k_i,i\}$, thus still leading to a covering of at least four nodes. Therefore, at the end of any block $b\geq 4$, each node will at least cover what had been covered by certain four nodes at the end of block $b-4$.

Inductively, it is easy to see that at the end of any block $b\geq 2^{m-2}$,  each node will at least cover what had been covered by certain $m$ nodes at the end of block $b-2^{m-2}$. Since each node covers itself by the end of block 0, for a network of any finite $n$ nodes, each node will cover the whole network, i.e., be able to decode some messages of any of the other nodes, at least by the end of block $b = 2^{n-2}$.

Before we conclude the proof, we need to demonstrate that the decoding delay is finite, so that there is no rate loss by block Markov coding. We use a contradiction argument. Suppose that the delay of some node $i$ decoding the messages of another node $j$ is not upper bounded, i.e.,
\begin{equation}
\label{eq2}
\limsup_{b\rightarrow \infty}[D_i(w_j(b))-b]=\infty
\end{equation}
where $D_i(w_j(b))$ denotes the block, by the end of which, node $i$ decodes the message $w_j(b)$---the block-$b$ message of node $j$. Since at the end of any block $b\geq 1$, node $i$ always decodes the block-$b$ transmission of another node, if \dref{eq2} holds, then there must exist another node $i_1\neq i$, such that
\begin{equation}
\label{eq3}
\limsup_{b\rightarrow \infty}[D_{i_1}(w_j(b))-b]=\infty.
\end{equation}
In fact, if no other nodes encounter an unbounded delay, then no other nodes will relay $w_j(b)$
with an unbounded delay, and then node $i$ will not decode $w_j(b)$ with an unbounded delay.

Now, since both $i$ and $i_1$ encounter unbounded delay in decoding $w_j(b)$, for the same reason as above, there must be a third node that encounters unbounded delay in decoding  $w_j(b)$. This argument can be continued, so that all the nodes have to encounter unbounded delay in decoding  $w_j(b)$, including node $j$ itself. This is obviously in contradiction. Therefore, \dref{eq2} cannot hold, i.e., all the decoding delays in the network must be uniformly bounded by some constant $T_0$.

\hfill ${\large \Box}$

Proofs of Theorems \ref{th2} and \ref{th3} follow similarly by treating every $K$ blocks together as a group block, and applying the argument above to the group blocks. Theorems \ref{th4} and \ref{th5} are simple applications.


\begin{thebibliography}{10}

\bibitem{xie07a}
L.-L. Xie, ``Network coding and random binning for multi-user channels,'' invited talk at the
  {\em 2007 IEEE Communication Theory Workshop}, (Sedona, Arizona),
  May 2007.

\bibitem{xie07}
L.-L. Xie, ``Network coding and random binning for multi-user channels,'' in
  {\em Proc. IEEE Canadian Workshop on Information Theory}, (Edmonton, Canada),
  June 2007.

\bibitem{xie08a}
L.-L. Xie, ``Omnidirectional relay in wireless networks,'' in {\em Proc. 2008
  {IEEE} International Symposium on Information Theory}, (Toronto, Canada),
  July 2008.

\bibitem{xie08b}
L.-L. Xie, ``Omnidirectional relay in wireless networks,'' submitted to {\em IEEE Trans. Inform. Theory}, November 2008.

\bibitem{covelg79}
T.~Cover and A.~{El Gamal}, ``Capacity theorems for the relay channel,'' {\em
  {IEEE} Trans. Inform. Theory}, vol.~25, pp.~572--584, 1979.

\bibitem{ahlcailiyeu00}
R.~Ahlswede, N.~Cai, S.-Y.~R. Li, and R.~W. Yeung, ``Network information
  flow,'' {\em IEEE Trans. Inform. Theory}, vol.~46, pp.~1204--1216, July 2000.

\end{thebibliography}
\end{document}